\documentstyle[jkas,natbib]{article}

\bibpunct[, ]{(}{)}{;}{a}{}{,}

\newcommand{\two}{{\it{SW-Field}}}
\newcommand{\love}{{\it{NE-Field}}}

\newcommand{\ind}[1]{${}^{#1}$}

\newcommand{\HII}{\ion{H}{2}\ }
\newcommand{\G}[1]{G#1--0.3}
\newcommand{\CO}{\ind{13}CO\ }
\newcommand{\ks}{$K_{\rm s}$}
\newcommand{\jk}{\mbox{$J$$-$\ks}}
\newcommand{\jh}{\mbox{$J$$-$$H$}}
\newcommand{\hk}{\mbox{$H$$-$\ks}}
\newcommand{\jhk}{$J$-, $H$-, and \ks-}

\newcommand{\ra}[3]{#1\ind{\mathrm{h}}#2\ind{\mathrm{m}}#3\ind{\mathrm{s}}}
\newcommand{\dec}[3]{#1\arcdeg#2\arcmin#3\arcsec}
\newcommand{\kmpers}{\mbox{km s\ind{-1}}}
\newcommand{\di}{\item[--]}

\beginpage{17}
\endpage{28}
\year{2007} \volume{40}

\runningauthor {H. KIM ET AL.} \year{2007} \volume{40}
\beginpage{17}\endpage{28}
\runningtitle{NIR STUDY OF W51B STAR-FORMING REGION}
\begin{document}

\title{A NEAR-INFRARED STUDY OF THE HIGHLY-OBSCURED ACTIVE STAR-FORMING REGION W51B}

\author{Hyosun Kim$^{1}$, Yasushi Nakajima$^{2}$, Hwankyung Sung$^{3}$, Dae-Sik Moon$^{4,5}$, and Bon-Chul Koo$^{1}$}
\address{$^1$ Astronomy Program, Department of Physics and Astronomy, Seoul National University, Seoul 151-742, Korea\\ {\it E-mail: hkim@astro.snu.ac.kr \& koo@astrohi.snu.ac.kr}}
\address{$^2$ National Astronomical Observatory of Japan, 2-21-1 Osawa, Mitaka, Tokyo 181-8858, Japan}
\address{$^3$ Department of Astronomy and Space Science, Sejong University, Gunja-dong 98, \\Gwangjin-gu, Seoul 143-747, Korea\\ {\it E-mail: sungh@sejong.ac.kr}}
\address{$^4$ Robert A.~Millikan Fellow, Division of Physics, Mathematics, and Astronomy, California Institute of Technology, MC 103-33, Pasadena, CA 91125\\ {\it E-mail: moon@srl.caltech.edu}}
\address{$^5$ Department of Astronomy and Astrophysics, University of Toronto, Toronto, ON M5S 3H4, Canada\\ {\it E-mail: moon@astro.utoronto.ca}}

\address{\normalsize{\it (Received January 26, 2007; Accepted March 20, 2007)}}

\offprints{H. Kim}

\abstract{We present wide-field $JHK_{\rm s}$-band photometric observations of the three compact
 \HII regions \G{48.9}, \G{49.0}, and \G{49.2} in the active star-forming region W51B.
 The star clusters inside the three compact \HII regions show the excess number of stars
 in the \jk~histograms compared with reference fields.
 While the mean color excess ratio ($E_{J-H}/E_{H-K_{\rm s}}$)
 of the three compact \HII regions are similar to $\sim$ 2.07,
 the visual extinctions toward them are somewhat different:
  $\sim$ 17 mag for \G{48.9} and \G{49.0}; $\sim$ 23 mag for \G{49.2}.
  Based on their sizes and brightnesses, we suggest that the age of each compact \HII region is $\le$ 2 Myr.
   The inferred total stellar mass, $\sim$ $1.4 \times 10^4 M_\odot$,
   of W51B makes it one of the most active star forming regions in the Galaxy with the star formation efficiency of $\sim$ 10 \%.}

 \maketitle

\section{INTRODUCTION}
W51B is a giant molecular cloud (GMC) complex located at the tangential point of the Sagittarius spiral arm of the Galaxy, and it is a very active star-forming region that consists of several compact \HII regions.
In radio continuum and molecular line images, the compact \HII regions appear to be still embedded in parental molecular clouds, indicating that they are relatively young \citep[e.g.][]{bieging, km97, koo97, koo99, mp98}. The \HII regions also show strong X-ray emission likely caused by magnetic activities of internal young stars \citep{koo02}. The line-of-sight velocities (61--70 \kmpers) of the sources in W51B are greater than the maximum kinematic velocity (54--58 \kmpers) permitted by the Galactic rotation \citep[and references therein]{koo99}. This has been known as the ``high-velocity stream'' in this region which might have resulted from the spiral density wave associated with the Sagittarius spiral arm.
The southern part of W51B appears to show an expanding ring structure with two molecular clumps associated with the compact \HII regions \G{48.9} and \G{49.0} \citep{mp98}.
Using 21-cm \ion{H}{1} absorption spectra toward the compact \HII regions, \citet{kolpak} determined the distance to W51B to be $\sim$ 5.6 kpc.\footnote{We will use 5.6 kpc as the distance to W51B in this paper.}

One of the most interesting features of W51B is its very high star formation efficiency (SFE): while the Galactic average value is 1--2 \% \citep{myers}, \citet{moon} obtained 16 \% for the overall SFE of W51B and \citet{koo99} did 7 \% and 15 \% for the north-eastern and south-western parts of W51B, respectively. Most of the previous studies of the star formation in W51B, including the ones carried out by \citet{moon} and \citet{koo99}, have exclusively used radio contiuum and molecular line data.
Recently, \citet{kumar} have obtained several small-field (= 90\arcsec) near-infrared (NIR) images of W51B, and identified a new, small-scale clumpy structure of star clusters of the compact \HII regions. However, the observations of \citet{kumar} were somewhat shallow, only reaching $\sim$16.2, $\sim$14.4, and $\sim$15.8 mag for $J$-, $H$-, and $K$-band, respectively. This rather shallow exposure, together with their small field of view, made it difficult to study fainter, extended stellar populations in W51B.
They calculated the mass of each cluster assuming an average visual extinction of 7 mag. With such moderate extinction, however, all of OB-type members should be visible in second Palomar Observatory Sky Survey (POSS II) red image \citep[$R_{c, \rm lim}$ = 20.8 mag,][]{reid}, while no optical counterpart is known. This implies therefore the extinction toward W51B \HII regions must be larger than 7 mag.
In this paper, we obtained deeper NIR photometric data that cover much more extended regions of W51B. With these new data, we find the initial mass function and the total stellar mass of star clusters in W51B, as well as NIR-based SFE of W51B. Our new observations confirm the results of the previous radio observations that W51B is one of the most active star forming regions in the Galaxy.

\section{OBSERVATIONS} 
\begin{figure}[!tb]
\epsfxsize=8.3cm \epsfbox{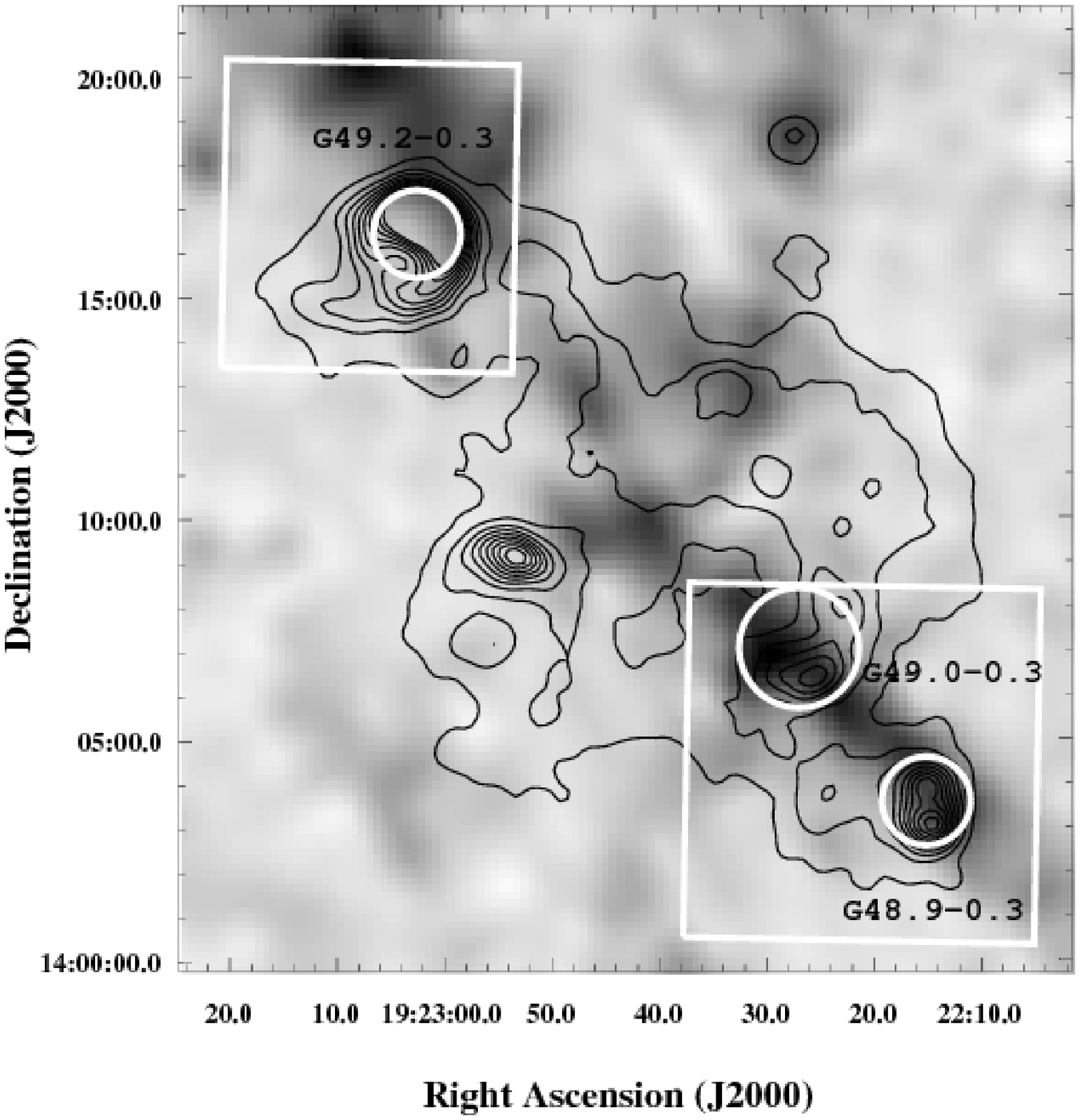} \caption{Gray-scale map of the
\CO J=1--0 line emission integrated over $v_{_{\mathrm{LSR}}}$
between +45 and +75 \kmpers~\citep{fcrao}, superposed on the black
contours of the 21-cm radio continuum map \citep{km97}. The
contour levels are 0.1, 0.2, 0.3, 0.4, 0.5, 0.6, 0.7, 0.8, 0.9,
and 1.0 Jy beam\ind{-1}. The white boxes and circles represent the
NIR observation fields and the NIR (especially \ks-band) nebulous
area (see Figure \ref{reg}): \two~including \G{48.9} and \G{49.0}
(from SW to NE), and \love~including \G{49.2}. The peak positions
of the radio continuum are consistent with those of the NIR
nebulosity.}\label{13CO_H21}
\end{figure}
\begin{figure}
\epsfxsize=8.3cm \epsfbox{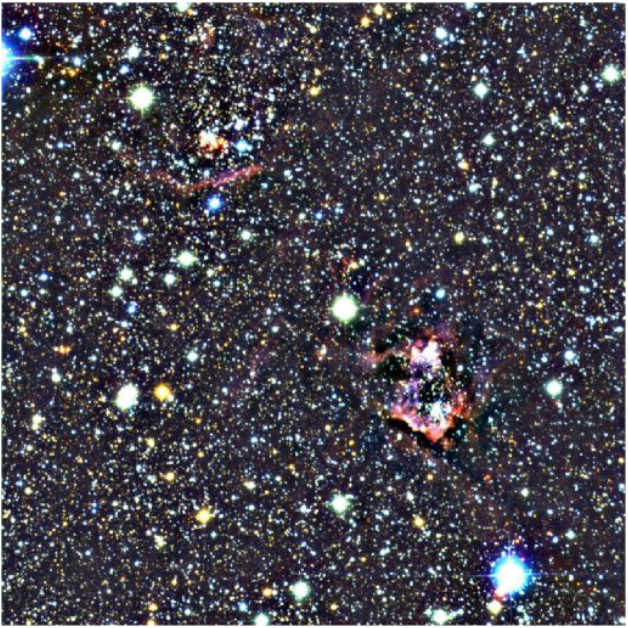}
\caption{$J$, $H$, and \ks\ three-color image of \two, including the two compact \HII regions \G{48.9} and \G{49.0}. The showing field is 8.2\arcmin$\times$8.2\arcmin~centered at $(\alpha_{J2000.0},\ \delta_{J2000.0})$ = (\ra{19}{22}{21}, \dec{+14}{4}{26}). The \jhk band data are represented as blue, green, and red, respectively.}\label{TwoNebula}
\end{figure}
\begin{figure}
\epsfxsize=8.3cm \epsfbox{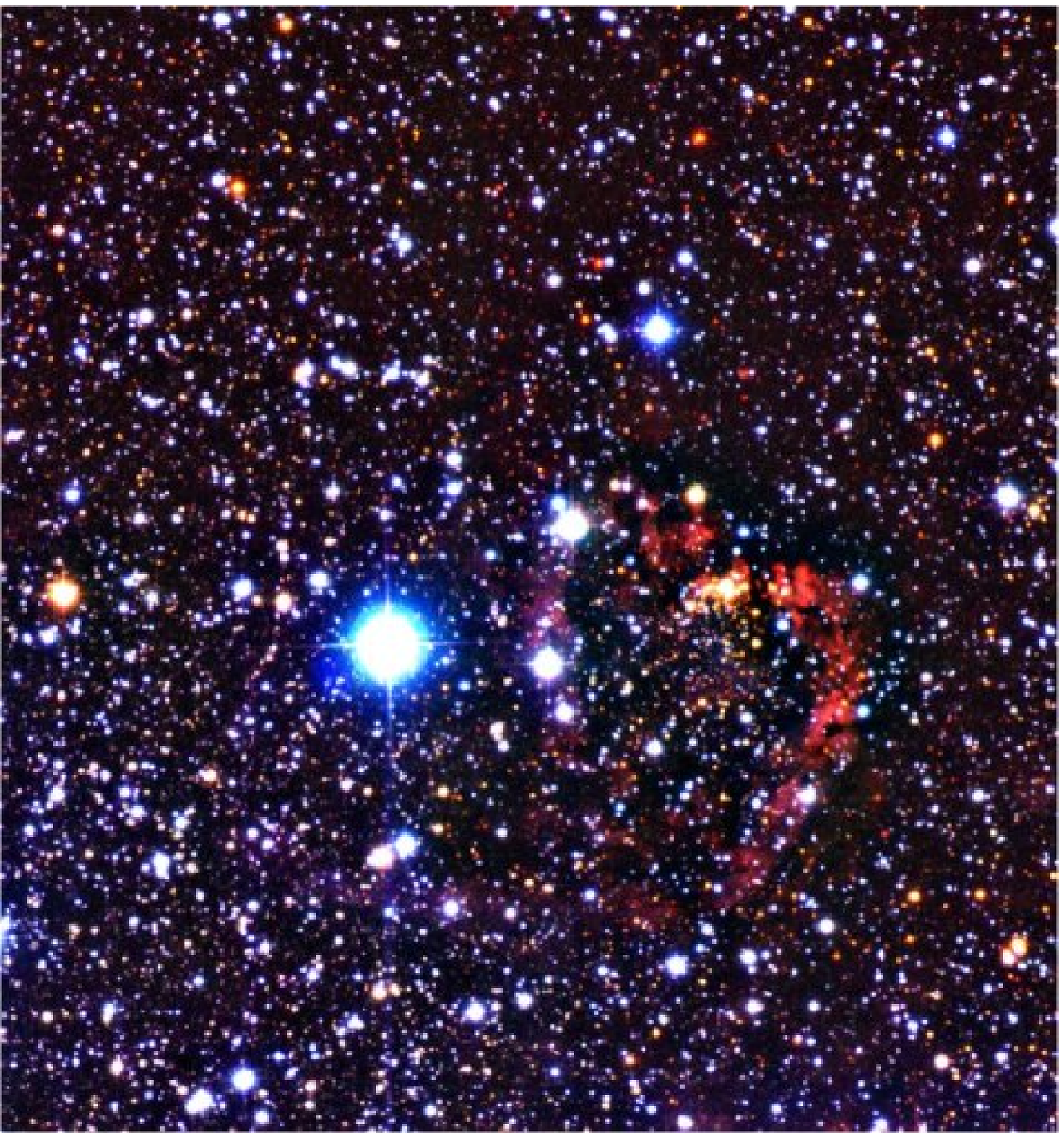}
\caption{Same as Figure \ref{TwoNebula} but for \love, which includes \G{49.2}. The showing field is 6.7\arcmin$\times$7.0\arcmin~centered at $(\alpha_{J2000.0},\ \delta_{J2000.0})$ = (\ra{19}{23}{7}, \dec{+14}{16}{47}).}\label{LoveNebula}
\end{figure}

We carried out near-IR $JHK_{\rm s}$-band photometric observations
of W51B using Wide-field InfraRed Camera (WIRC) aboard the 5-m
Palomar Hale telescope on 2003 June 17 and 18. WIRC is equipped
with a HAWAII-II HgCdTe 2K array and the pixel scale is
0.25\arcsec~pixel\ind{-1}, providing a 8.7\arcmin $\times$
8.7\arcmin\ field of view. The typical seeing was 1\arcsec\ over
the observations. We selected two fields based on the previous
radio observations, and Figure \ref{13CO_H21} shows the locations
of the observed fields superimposed on previous radio images. As
in Figure \ref{13CO_H21}, one field includes \G{48.9} and
\G{49.0}, while the other field contains \G{49.2}, the brightest
radio continuum source in W51B \citep{km97}, at its center. We
shall call the former \two\ and the latter \love\ hereafter. We
obtained 30 and 60 dithered frames of \two\ and \love\,
respectively, for each $JHK_{\rm s}$ band.  The exposure of
individual frame was 10 sec. Each dithered frame was shifted and
combined after subtracting dark and median-combined sky frames and
correcting flat field. Figures \ref{TwoNebula} and
\ref{LoveNebula} present the final three-color images of the two
fields, where we can identify extended emission around the \HII
regions due to Br-$\gamma$ emission in the \ks\ band.

In order to find stars in our WIRC images, we iteratively used the
PSF-fitting program in IRAF\footnote{IRAF (Imaging Reduction \&
Analysis Facility) is distributed by the National Optical
Astronomy Observatories, which are operated by the Association of
Universities for Research in Astronomy, Inc., under cooperative
agreement with the National Science
Foundation.}/D\-A\-O\-P\-H\-O\-T until the residuals show no
apparent stars left in the images. As a result, we have detected
$\ge$ 11000 and $\ge$ 7000 stars in the \two\ and \love,
respectively. Among them $\sim$ 10000 and $\sim$ 6000 stars have
signal-to-noise ratio greater than 10 in all the $JHK_{\rm s}$
bands, within 10-$\sigma$ limiting magnitudes of $J$ = 19.8, $H$ =
19.0, and \ks\ = 17.9 mag. For photometric and astrometric
solutions, we used $\sim$ 200 stars of 2MASS Point Source
Catalog\footnote{http://www.ipac.caltech.edu/2mass/} in our
fields. The magnitudes of the faintest stars were $J$ = 21.5 $\pm$
0.5, $H$ = 19.7 $\pm$ 0.2, and \ks\ = 19.2 $\pm$ 0.4 for \two\ and
$J$ = 21.5 $\pm$ 0.5, $H$ = 19.8 $\pm$ 0.2, and \ks\ = 19.5 $\pm$
0.5 for \love\ respectively, while the magnitude of the brightest
and unsaturated star was $\sim$ 10 in all bands. Our astrometric
solutions gave 1-$\sigma$ uncertainties of $\sim$0\farcs3 for
\two\ and $\sim$0\farcs1 for \love\ with respect to the 2MASS
system.

\section{Star Clusters in W51B} \label{sec:cluster}
\subsection{Membership Selection} 
\begin{figure*}
\epsfxsize=17.5cm \epsfbox{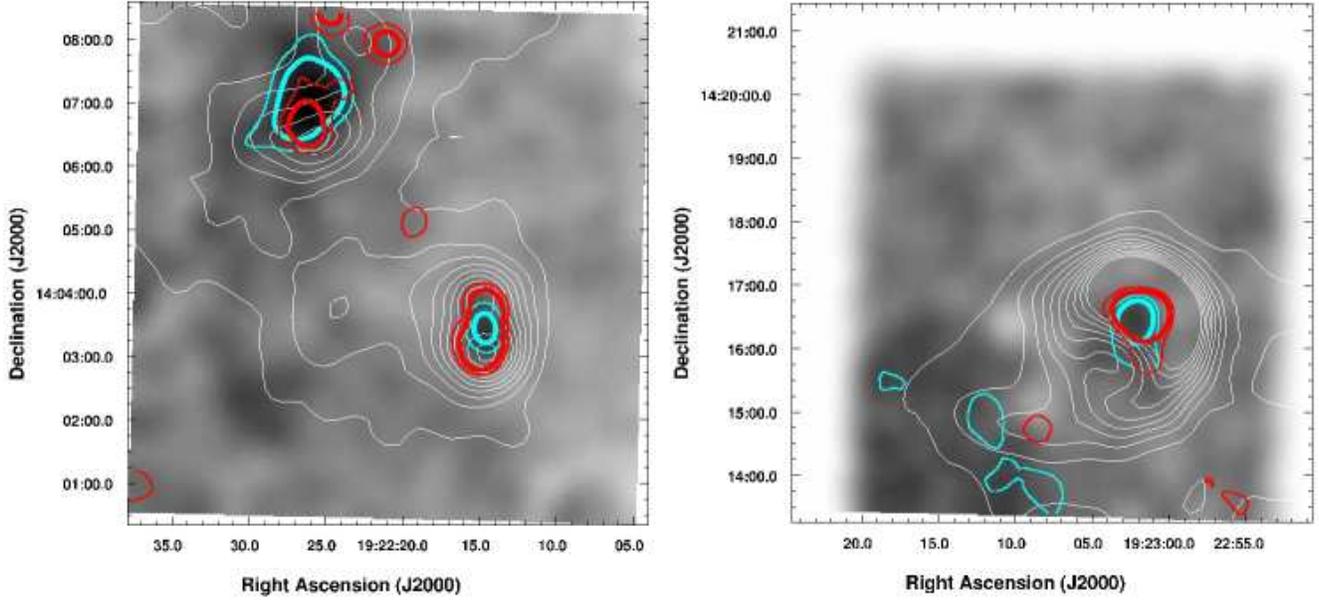} \caption{Gray-scale map of the
number density of stars detected in all bands, superposed on red
and blue contours of the number densities of ``P'' and ``B''
members, respectively (see text), for each field. The colored
thick and thin contours represent 3-$\sigma$ and 2-$\sigma$ of the
number density where $\sigma$ is the standard deviation of each
kind of density. The white contours are the 21-cm radio continuum
whose levels are given in Figure \ref{13CO_H21}. The left and
right panels are for \two~and \love,
respectively.}\label{contours}
\end{figure*}

\begin{figure*}
\epsfxsize=17cm \epsfbox{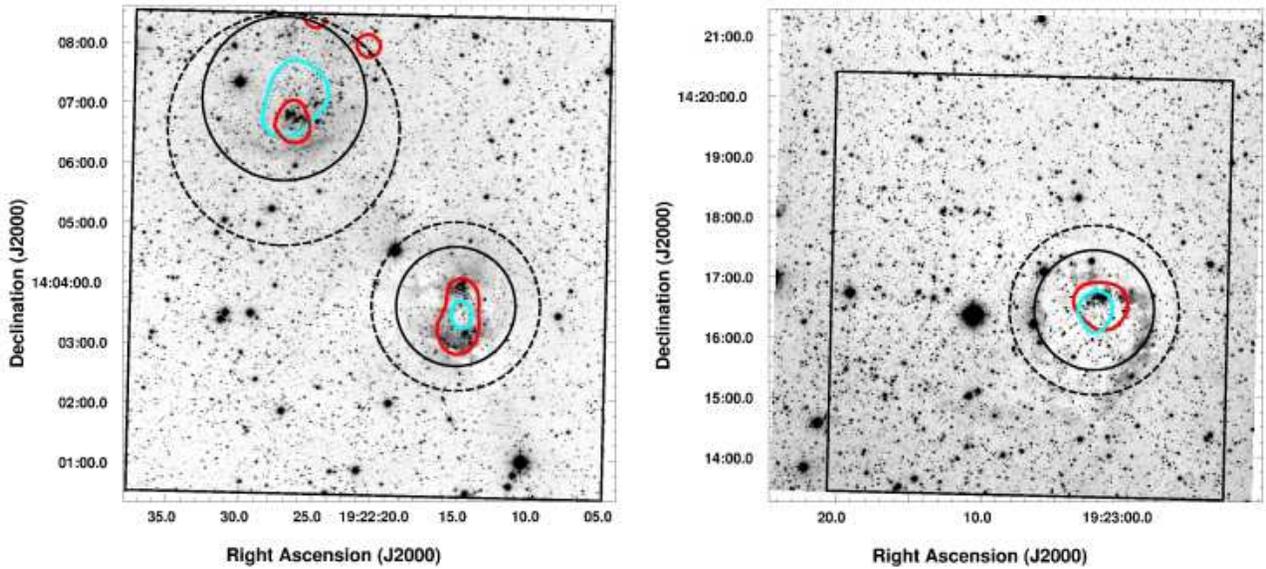}
\caption{\ks-band images for \two~(left) and \love~(right). For \love, the observed \jhk band frames were shifted each other, and we use only the matched area which is inside the black box. The solid and dashed circles represent the cluster and reference regions, respectively. The reference region has the same area with the corresponding cluster region. Also shown are the cluster-core regions which are defined inside the 3-$\sigma$ stellar density contours in Figure \ref{contours}.}\label{reg}
\end{figure*}

In order to investigate the existence of noticeable star clusters
inside the \HII regions, we first examine the stellar number
density map as in Figure~\ref{contours} which was obtained by
Gaussian smoothing of number of stars with 15\arcsec\ FWHM. It
shows clusterings of stars inside the three compact \HII regions.
Although there are many stars at south-east of \love, they should
not be regarded as cluster members because they are randomly
distributed outside the cloud. They may be background stars behind
W51B GMCs. In order to determine the cluster members and examine
their properties, we define the cluster regions and the
corresponding reference regions on the basis of the positions of
\HII regions, which are described as solid and dashed circles,
respectively, in Figure \ref{reg}. Each reference region is
selected just outside the cluster region to minimize the offset
effect of spatial density fluctuation. A reference region has the
same area with the corresponding cluster region. There are about
600, 1500, and 500 stars in the cluster regions within the \HII
regions \G{48.9}, \G{49.0}, and \G{49.2}, respectively. These
numbers of stars inside the cluster regions are at least
1-$\sigma$ larger than the counts in the randomly selected regions
having same area. For \G{49.2}, we avoid selecting the random
regions outside molecular cloud because there must be a lot of
background field stars in the off-cloud area.

\begin{figure}
\epsfxsize=9cm \epsfbox{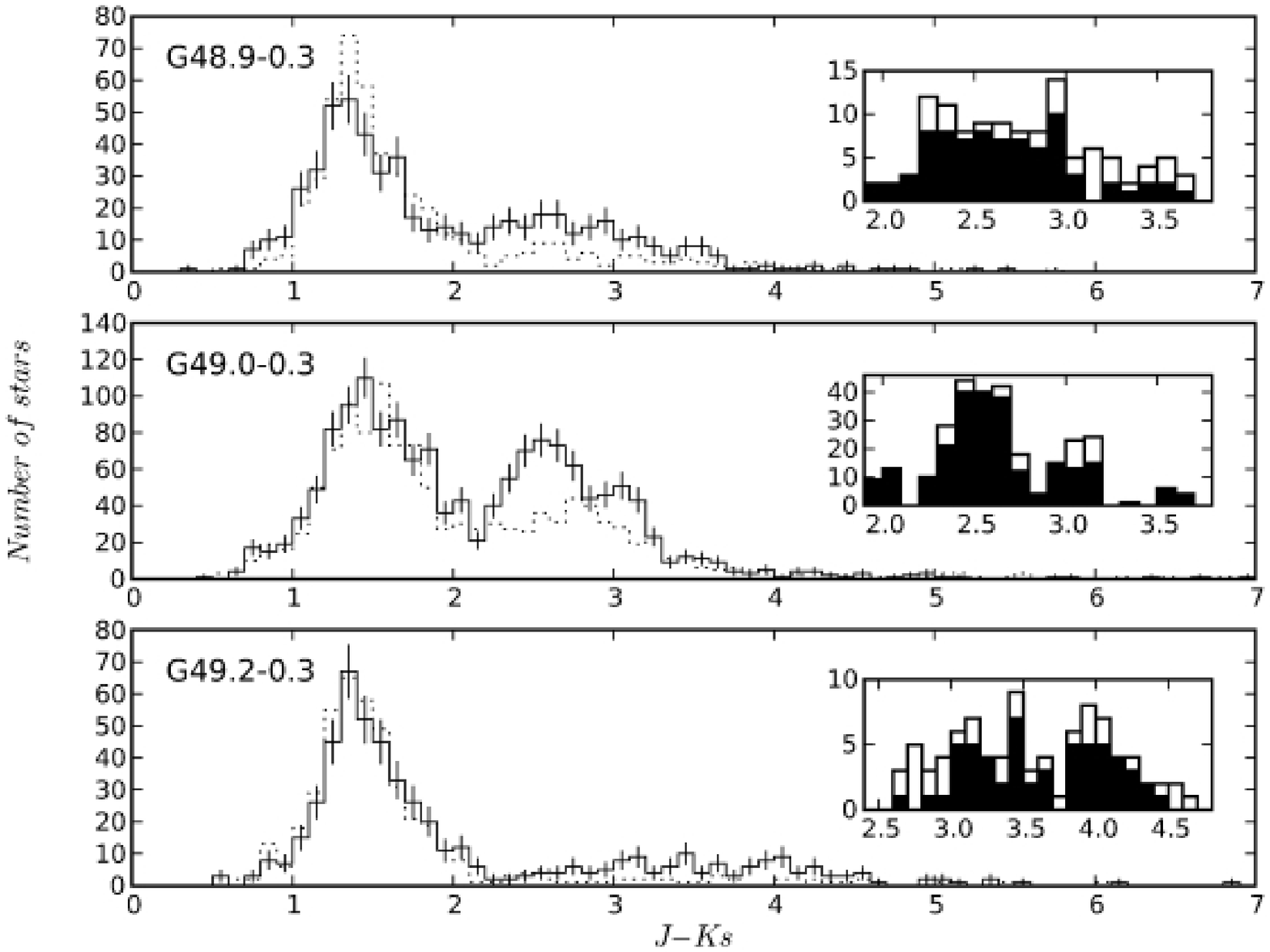} \caption{Histograms along
\jk~color. Solid and dotted lines show the histograms for a
cluster region and the corresponding reference region,
respectively. The error bar denotes the Poisson noise level in
each bin. The difference for the second bump in the number of
stars between the cluster region and the reference region is
displayed as an inset of each panel. The white and black
histograms represent the numbers of PMS and the others,
respectively.}\label{Avhist}
\end{figure}

Figure \ref{Avhist} shows the histograms of \jk~color of the stars
detected at all bands in the direction of three \HII regions. The
first bump at \jk~$\sim$ 1.4 is found on the histograms for both
cluster and reference regions, thus it should be caused by
foreground field stars. With an assumption of $A_V$ $\sim$ 1 mag
$\mathrm{kpc}^{-1}$, they seem to be mainly composed of the
Sagittarius arm which extends through a range of $\sim$ 3--6 kpc
at $l \simeq 49\arcdeg$ \citep{wainscoat}. We see outstanding
bumps toward \G{48.9} and \G{49.0} at 2.2 $<$ \jk~$<$ 3.2,
compared to the histograms of the reference regions. In the
histogram for \G{49.2}, the extra bump is at 2.6 $<$ \jk~$<$ 4.5.
The inset panel represents the excess number of stars compared to
reference region. The excess should be caused by the probable
cluster members.


\begin{figure}
\epsfxsize=8.8cm \epsfbox{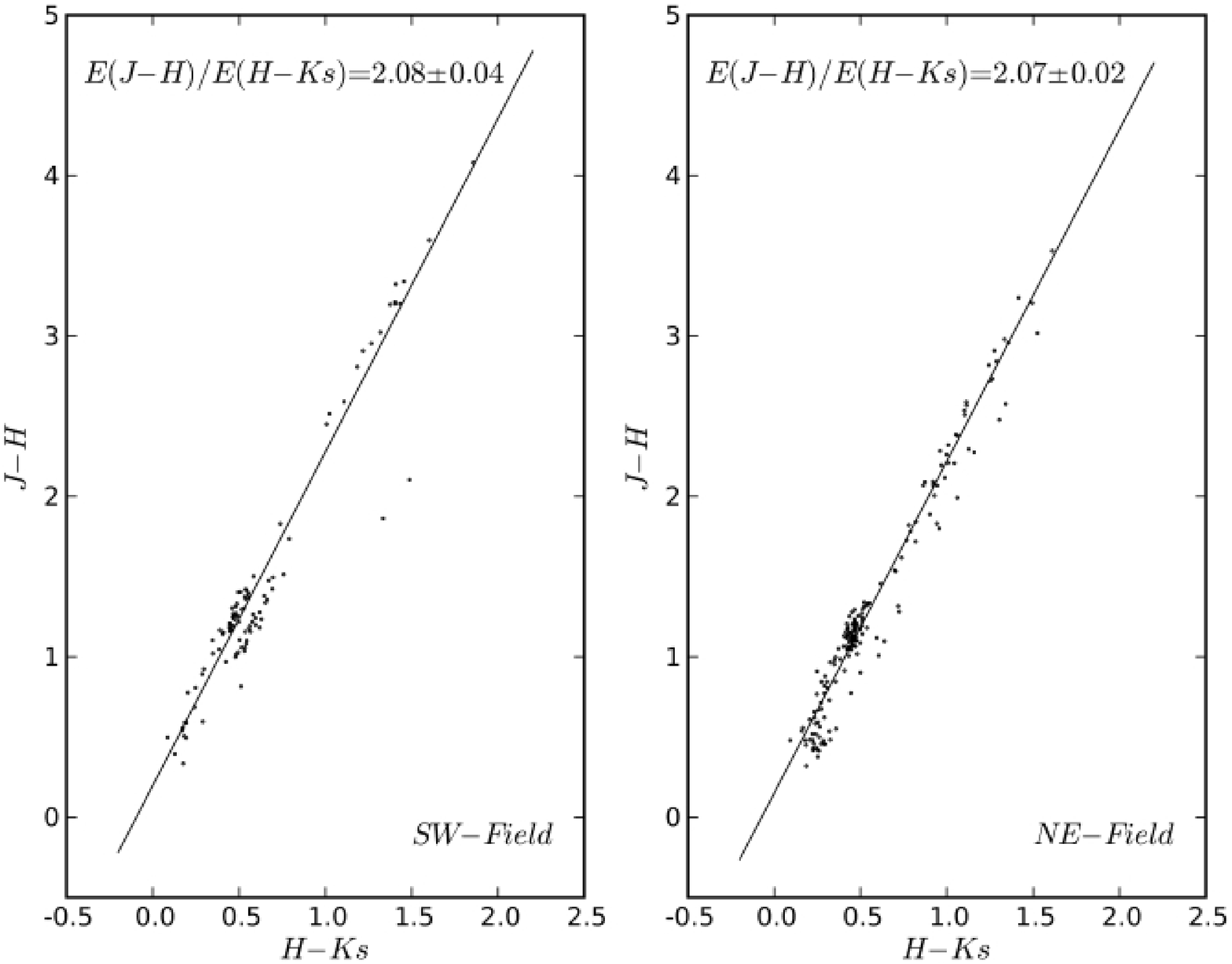} \caption{Color-color diagram of
the bright stars (\ks$<$13) in each field with good photometry
(\hk~and \jh~color errors less than 0.01 mag). Reddening slopes
are obtained using a least-squares fit with 3-$\sigma$ clipping.
The left panel is for \two, and the right one for \love. The line
in each panel shows the fitted reddening slope, both of which are
$\sim$ 2.07.}\label{ccmr}
\end{figure}

Although the bump at \jk~$\ge$ 2.2 probably includes most of the
cluster members (we call them ``B'' (Bump) members hereafter), we
may fail to select the pre-main-sequence (PMS) stars which are
generally redder than the main-sequence (MS) stars due to the
intrinsic emission from the circumstellar dust. Thus, we need to
separately include the PMS stars into the cluster member
candidates. We use \hk~vs. ~\jh~color-color diagram (CCD) to select
PMS stars with intrinsic NIR color excess. In order to distinguish
the intrinsic color excess from the reddening by interstellar
extinction, we first compute the color excess ratio
$E_{J-H}/E_{H-K_{\rm s}}$ toward W51B in the 2MASS system by a
least-squares fit with 3-$\sigma$ clipping. Figure \ref{ccmr}
shows the CCD of bright stars (\ks~$<$ 13) with each color error
less than 0.01 mag. The stars after 3-$\sigma$ clipping yield the
ratios of $2.08 \pm 0.04$ for \two~and $2.07 \pm 0.02$ for \love.
Those are similar but slightly larger than 1.98 of \citet{fitz}
provided from extensive data in the Galaxy or $1.97 \pm 0.01$ of
\citet{han} obtained for a 25\arcmin$\times$25\arcmin~field
including whole W51B giant molecular cloud region and large
off-cloud region. Because our fields are focused on the cloud core
regions, there could be a little difference from the value
calculated for a wide field. Hereafter, we apply the color excess
ratio of 2.07 in the direction of the \HII regions in W51B and use
the relations of \citeauthor{fitz} in order to estimate total
extinction.

\begin{figure*}
\epsfxsize=17.5cm \epsfbox{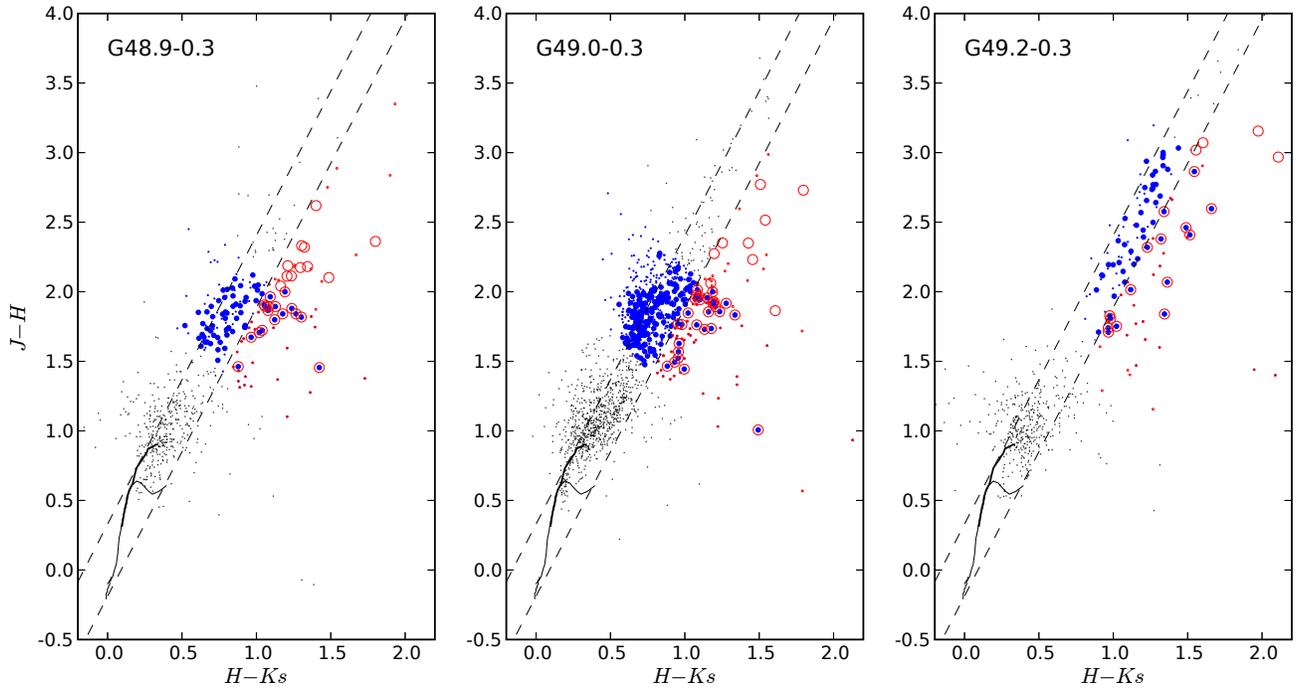} \caption{Color-color diagram of
stars in each cluster region. The member candidates are marked in
red and blue colors. The red marks are ``P'' members, while the
blue marks are the ``B'' members. The red open circles and the
blue closed circles represent the member candidates with small
color error ($<$ 0.1 mag), and the colored dots are the candidates
with larger error. The thin and thick solid curves are the loci
occupied by the unreddened dwarfs and giants \citep{BB, koorn}.
The dashed lines are parallel to the reddening
vector.}\label{ccds}
\end{figure*}

\begin{figure*}
\epsfxsize=17.5cm \epsfbox{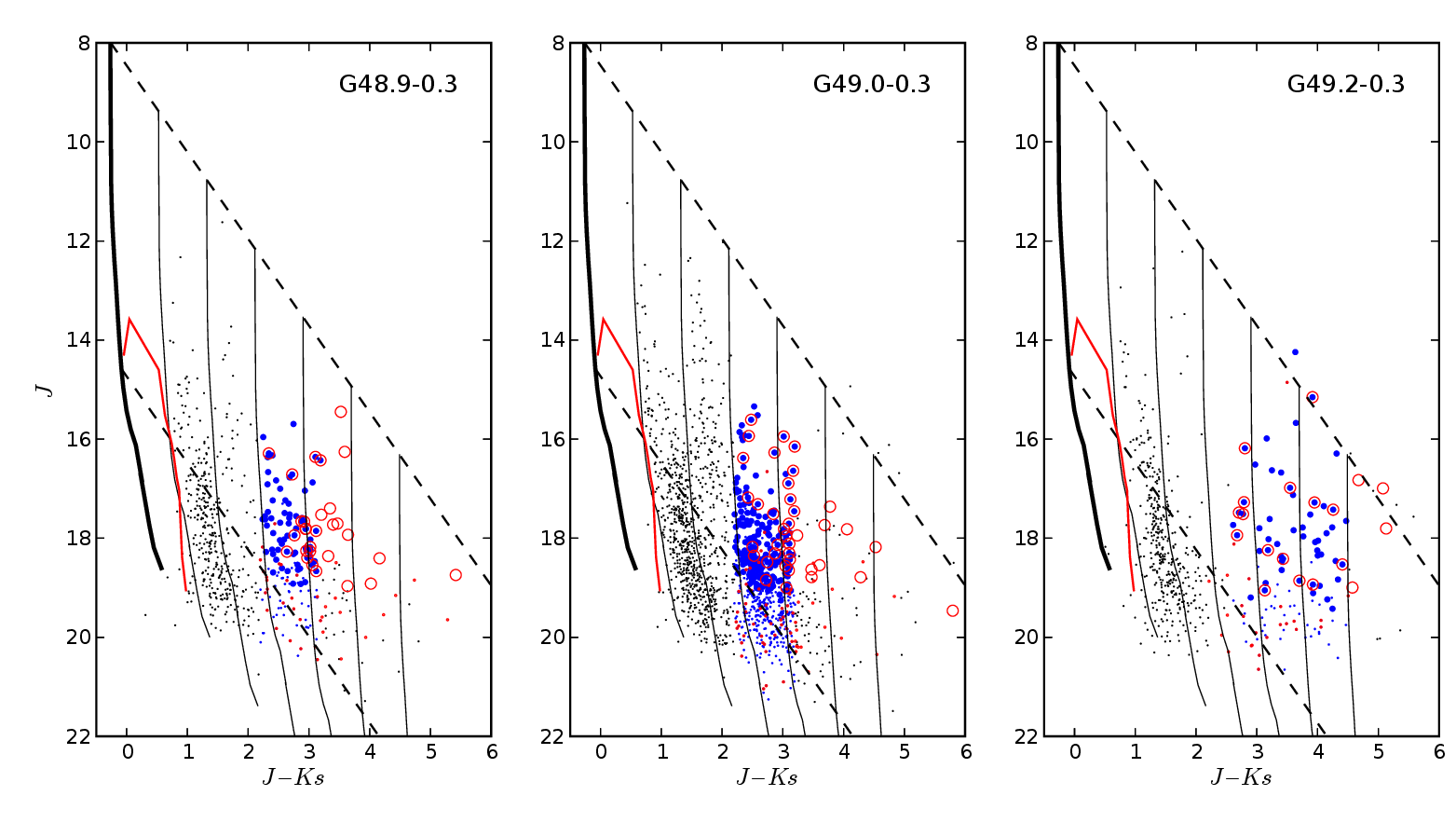} \caption{Color-magnitude diagram
of stars in each cluster region. The black, thick curve represents
the unreddened zero-age main-sequence (ZAMS) relation for Z=0.020
from the Geneva models \citep{schaller, bessell98}, adopting the
distance of 5.6 kpc to the compact \HII regions. The dashed lines
are parallel to the reddening vector of ZAMS star having mass of
120$M_\odot$ (up) and 3$M_\odot$ (down). The black, thin curves
are reddened ZAMS corresponding with $A_V$ of 5, 10, 15, 20, 25,
and 30 mag. Additionally, the red curve is the loci of unreddened
PMS stars with age of 1Myr \citep{testi}. The red and blue marks
represent the member candidates which are described in Figure
\ref{ccds}.}\label{cmds}
\end{figure*}

Figure \ref{ccds} is \hk~vs.~\jh~CCDs where the reddening slope of
the dashed lines is determined by the color excess ratio obtained
in the previous paragraph. The area between the two dashed lines
is the ``reddening zone'' in which the reddened dwarfs and giants
fall. Stars with intrinsic infrared excess emission can fall into
the right of the reddening zone. \citet{ladanadams} called this
region as ``excess zone''. The sources in the excess zone are
generally pre- and post-main-sequence stars with circumstellar
dust and galaxies with redshifts z $>$ 0.1. Toward the compact
\HII regions, most of them should be pre-main-sequence stars. We
mark these stars in red color, except possible spreading
foreground stars with \jk~$<$ 2.2. We call them ``P'' (PMS)
members hereafter. And we mark the ``B'' members in blue color.
Most of the ``B'' members are MS stars, though some of them are
marked in both colors which indicate that they are probably PMS
stars. Since most stars in these categories are located above the
reddening line of $3 M_\odot$ stars in \jk~vs.~$J$ color-magnitude
diagrams in Figure \ref{cmds}, we suppose the observed cluster
members to be OB-type MS stars and/or intermediate- to high-mass
PMS stars. We note the numbers of stars in the two categories in
Table \ref{group}. The selected cluster member candidates are 168,
589, and 120 for \G{48.9}, \G{49.0}, and \G{49.2}, respectively.
These candidates, however, include some field star contamination
though we already excluded most of the foreground field stars
composing the bump at \jk~$<$ 2.2 in Figure \ref{Avhist}.

\begin{deluxetable}{cccc}
  \tablewidth{0pt}
  \tablecaption{Number of Cluster Member Candidates  \label{group}}
  \tablehead{\colhead{\HII region} & \colhead{``B'' members}
    & \colhead{``P'' members} & \colhead{``B'' \& ``P'' members}}
  \startdata
    \G{48.9}   & 143 & \phm{1}66 & 41 \\
    \G{49.0}   & 560 &       105 & 76 \\
    \G{49.2}   & 109 & \phm{1}45 & 34 \\
  \enddata
  \tablecomments{ The fourth column represents the number of stars in both categories, and we include these stars into PMS member candidates. MS member candidates are considered as the stars in ``B'' but not ``P'' category.}
\end{deluxetable}

In order to analyze how many field stars are contained, we choose out the stars inside the reference region in each category using the same method, and compare the number of stars between cluster and reference regions.
We confirm that at least 60--80 \% of the selected member candidates are statistically reliable for \G{48.9} and \G{49.2}, while the reliability is 40 \% for \G{49.0}. Since \G{49.0} is sparsely distributed and the extinction varies a lot, the cluster members could not be determined more reliably.
However, most of the field star contamination contributes to the low luminosity regime ($J > 18$ mag), thus their contribution to the total luminosity is expected to be insignificant.

\subsection{Interstellar Extinction toward W51B Clusters} \label{ssec:ext}
We expect that the average interstellar extinction toward the
cluster can be simply obtained from the mean photometric color of
members. Since the range of the intrinsic NIR colors of OB-type MS
stars is very narrow ($\Delta (J-K_{\rm s})_{\mathrm{intrinsic}}
\sim$ 0.2 mag) compared to the \jk~range of sources in the
observed field, the \jk~color can be a good indicator of
extinction.

We transform the observed \jk~colors of MS member candidates to the visual extinctions using
\begin{equation}
  A_V=6.31\ [(J-K_{\rm s})_{\mathrm{obs.}}-(J-K_{\rm s})_{\mathrm{intrinsic}}],
\end{equation}
which is obtained from \citeauthor{fitz}'s relations converted to
the 2MASS system\footnote{http://www.ipac.caltech.edu/2mass/}
which is based on \citet{carpenter}. We assume the intrinsic
\jk~color of OB stars to be approximately zero. PMS stars must
possess redder intrinsic colors though there is not a definite
model explaining the color of massive PMS stars yet. We assume the
average extinction of PMS stars is identical to that of MS stars.
The mean visual extinctions for the member candidates are 17 mag
for both \G{48.9} and \G{49.0}, and 23 mag for \G{49.2}.
Similarly, mean $J$-band extinctions are obtained from $A_J=1.73\
(J-K_{\rm s})_{\mathrm{obs.}}$: 4.6 mag for \G{48.9} and \G{49.0},
and 6.2 mag for \G{49.2}. The photometric diagrams of \G{49.2}
show a wide separation between the cluster and foreground stars at
\jk~$\sim$ 2, which is consistent with the higher extinction value
than the other two clusters.

\subsection{Spatial Distribution of Member Stars} 
In order to see the clustering feature of the member candidates rigorously, we describe the number densities of the ``P'' and ``B'' members as red and blue contours in Figure \ref{contours}, overlapped on the gray-scale map for the number density of all observed stars. We define the cluster-core region inside the contour of 3-$\sigma$ where $\sigma$ is the standard deviation of the density fluctuations.
In Table \ref{number}, we display the stellar column and volume densities of each group inside each cluster-core region after statistical subtraction of the stars outside the region under consideration. They are the lower limits of the cluster densities because the lowest mass of the marginally detectable members are 2--5$M_\odot$ corresponding to the 10-$\sigma$ limiting magnitudes.

\begin{deluxetable}{c c r@{ $\pm$ }l c r@{ $\pm$ }l r@{ $\pm$ }l}
  \tablewidth{0pt}
  \setlength{\tabcolsep}{4mm}
  \tablecaption{Number of Stars in Each Cluster-Core Region \label{number}}
  \tablehead{\colhead{\HII region} & \colhead{Name \tablenotemark{a}}
    & \multicolumn{2}{c}{Number} & \colhead{Area (\sq\arcsec)}
    & \multicolumn{2}{c}{$N/A_{\mathrm{pc}^2}$ \tablenotemark{b}}
    & \multicolumn{2}{c}{$N/V_{\mathrm{pc}^3}$ \tablenotemark{c}}}
  \tablecolumns{9}
  \startdata
    \G{48.9}
      & P4855044--0017085 &  36&7  & 2450       &  63&12 & 20&4 \\
      & B4855164--0016594 &  23&6  & \phm{2}449 & 218&55 & 160&40 \\
    \tableline
    \G{49.0}
      & P4859241--0018016 &   8&4  & 1125       &  30&16 & 14&7 \\
      & B4859462--0017400 & 122&16 & 3674       & 142&18 & 36&5 \\
    \tableline
    \G{49.2}
      & P4912136--0020508 &  16&5  & 1989       &  34&10 & 12&4 \\
      & B4912111--0021025 &  26&7  & 1064       & 105&27 & 50&13 \\
  \enddata
  \tablenotetext{a}{\ The position of the local peak in Galactic coordinates were used to name the clustering groups. The first letter represents ``PMS'' and ``Bump'', respectively.}
  \tablenotetext{b}{\ Column density. $N/A_{\mathrm{pc}^2} = 4262\ N/A_{\sq\arcsec}$~ adopting the distance of 5.6 kpc.}
  \tablenotetext{c}{\ Volume density. $N/V_{\mathrm{pc}^3} = 66430\ N/A^{3/2}_{\sq\arcsec}$~ if we assume the cluster is spherical.}
\end{deluxetable}

In \G{48.9}, both MS and PMS members are well confined inside the area of radio continuum, elongated from north to south. The PMS members appear to be peanut-shaped.
\G{49.0} shows a large and bright density peak of ``B'' members. The PMS stars are concentrated on the southern part of it near the radio continuum peak. There are two small PMS clusters at rather distant places ($\sim$ 2\arcmin) on the north-west of \G{49.0}.
The central region of \G{49.2} is crowded inside a small place, while the radio continuum source is much larger than the previously mentioned two other \HII regions. If \G{49.2} is a ``blister-type'' \HII region being suggested from the radio continuum morphology and the distribution of molecular gas \citep[e.g.][]{koo99}, the large size of the ionized gas could be explained by low density of ambient medium.

\section{STAR FORMATION EFFICIENCY} 
\begin{figure*}
\epsfxsize=17.5cm \epsfbox{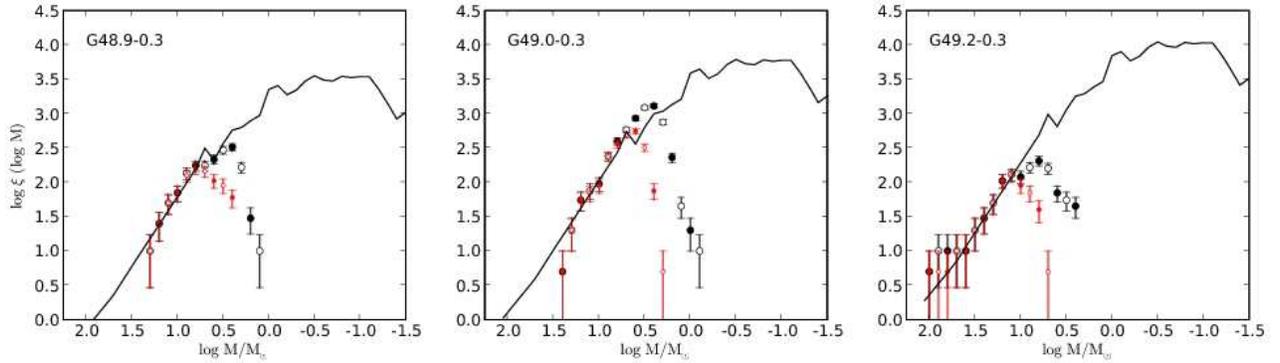} \caption{Initial mass functions
for all cluster member candidates (black) and the member
candidates with color error less than 0.1 mag (red). We count the
stars within the mass interval $\Delta \log\,(M/M_\odot) = 0.2$
and in the same-sized bin shifted by 0.1 to reduce the binning
effect. They are represented as open and closed circles. The error
bars denote the statistical fluctuations in the counts under the
assumption that the Poisson statistics governs the fluctuations.
The initial mass function of NGC 2264 (solid curve) is vertically
shifted to fit our data and used to estimate total mass of
W51B.}\label{imfs}
\end{figure*}

With the $J$-band magnitudes of member stars, we find the initial mass function (IMF) and the total stellar mass of each cluster. We use $J$-band luminosity which is less affected by the intrinsic excess emission of PMS stars than $H$- or \ks-band luminosity.
We transform the observed $J$ magnitudes of all member candidates to the absolute magnitudes applying the mean extinction of the member candidates derived at \S~\ref{sec:cluster}~\ref{ssec:ext}. The transformation from the absolute $J$-magnitude to the initial mass is accomplished using the theoretical mass--luminosity relation from the Geneva models and theoretical colors of MS stars \citep{schaller, bessell98} and \citet{testi} for PMS stars less massive than $3 M_\odot$. Since massive PMS stars ($> 3 M_\odot$) would be located very close to the main sequence in the mass--luminosity relation, the MS relation is applied for them. We assume the age of clusters to be 1 Myr, and the derived total mass is not sensitive to the age.
Figure \ref{imfs} shows the resultant IMF for each cluster.
The number of stars increases as the stellar mass decreases at the high mass regime, while the detection limit causes the IMFs to decrease at the lower mass regime.
For comparison, we plot the IMF of the young open cluster NGC 2264 \citep[solid curve]{sung} extrapolated using the IMF of the solar neighborhoods \citep{lee} for the massive stars ($> 30 M_\odot$) and vertically shifted to fit our data within $1.0 \leq \log\,(M/M_\odot) \leq 1.3$. \citeauthor{sung} noticed that NGC 2264 has similar shape of IMF to the Pleiades and the Trapezium.
The IMFs of \G{48.9} and \G{49.0} agree well with the comparison IMF at $M >$ 5--6 $M_\odot$. For \G{49.2}, the consistent range is $M > 10 M_\odot$ due to high extinction toward this cluster.

In order to estimate total mass of a cluster, the universality of IMF is assumed.
 The total mass is determined from
\begin{equation}
  \mathrm{M}_{\mathrm{tot}} = \int M \xi\,(\log\,M)\ d\,\log\,M
\end{equation}
where $\xi$ is the IMF of NGC 2264 which described in the previous paragraph. T
he upper limit of the integral is decided on the mass of the most massive stars, and the lower limit is fixed to $0.01 M_\odot$. Table \ref{mass} represents the total mass of the cluster associated with each \HII region. We also list the stellar mass of the only observed stars, instead assuming the universality of IMF. Undetected stars would contribute to the cluster mass in amount of at least twice than the observed massive stars do.

\begin{deluxetable}{cccc}
  \tablewidth{0pt}
  \setlength{\tabcolsep}{3mm}
  \tablecaption{Total Stellar Mass Estimations \label{mass}}
  \tablehead{\colhead{\HII region} & \colhead{Method} & \colhead{Mass Range}
    & \colhead{Total Stellar Mass} \\
    \colhead{} & \colhead{} & \colhead{($M_\odot$)} & \colhead{($M_\odot$)}}
  \startdata
    \G{48.9} & 1 &       0.01 $\sim$ 20 & 2400 \\
             & 2 & \phm{0}1.3 $\sim$ 20 & \phm{0}800 \\
    \tableline
    \G{49.0} & 1 &       0.01 $\sim$ 25 & 4200 \\
             & 2 & \phm{0}0.8 $\sim$ 25 & 2300 \\
    \tableline
    \G{49.2} & 1 &       0.01 $\sim$ 100 & 7800 \\
             & 2 & \phm{0}2.5 $\sim$ 100 & 1300 \\
  \enddata
  \tablecomments{ For the method (1), we assume the combined IMF of NGC 2264 and solar neighborhoods as a universal IMF in order to estimate total mass including undetected low-mass stars. Method (2) is the stellar mass of the observed stars without the assumption.}
\end{deluxetable}

We calculate the star formation efficiency (SFE) which is defined
as the ratio of the stellar mass to the total mass of stars plus
cloud, using total stellar mass in Table \ref{mass} and the
molecular cloud mass in Table 1 of \citet{koo99}. The overall SFEs
are 7 \% for the cloud SW including \G{48.9} and \G{49.0}, and 17
\% for the cloud NE including \G{49.2}. Even if we only use the
stellar mass of the observed massive stars, the efficiencies are
estimated to be 3--4 \%. Because no cluster is composed of only
massive OB stars, the SFEs of W51B appear to be clearly high
compared to the Galactic average value of 1--2 \%. Besides, the
high SFEs are lower limits because we ignore the possible smaller
clusters/associations along the GMCs. For explanations about the
high SFE, a cloud-cloud collision between high-velocity (68
\kmpers) cloud and the W51 GMC \citep{fcrao} and the enhancement
due to the spiral density wave \citep{koo99, kumar} have been
speculated.

Table 2 of \citet{ladanlada} shows that the SFEs for nearby embedded clusters range from $\sim$ 10 \% to 30 \% when the cloud core masses are adopted, not the masses for entire giant molecular clouds. The compressed core masses presented in that paper are less than 1000 $M_\odot$, which is at least 10 times less than the entire mass of W51B GMCs. We cannot compare the core SFEs to the overall SFEs of W51B GMCs until the individual core masses are obtained by high resolution observations.

\section{AGE OF CLUSTERS} 
\begin{deluxetable}{cccccc}
  \tablewidth{0pt}
  \setlength{\tabcolsep}{3mm}
  \tablecaption{Properties of the Clusters Associated with Three \HII regions
    \label{para}}
  \tablehead{
    \colhead{\HII region} & \colhead{$\langle A_V \rangle$} & \colhead{Radius}
    & \colhead{Spectral Type of} & \colhead{$\log N_C$ \tablenotemark{a}}
    & \colhead{Age \tablenotemark{b}} \\
    \colhead{} & \colhead{(mag)} & \colhead{(pc)}
    &\colhead{the most massive star} &\colhead{(s\ind{-1})}&\colhead{(Myr)}}
  \startdata
  \G{48.9} & 17 & 1.6 & B0 & 47.85 (46.90) & 0.6 (1.1) \\
  \G{49.0} & 17 & 2.2 & O9 & 48.31 (48.23) & 0.8 (0.9) \\
  \G{49.2} & 23 & 2.7 & O4 & 50.43 (50.25) & 0.3 (0.3) \\
  \enddata
  \tablenotetext{a}{\ Lyman continuum photon flux of all members (and the members without NIR excess).}
  \tablenotetext{b}{\ Dynamic age of the \HII region using all members (and the members without NIR excess).}
\end{deluxetable}

As the stars emit Lyman continuum photons, the \HII region increases in size. Therefore, in a given size of the \HII region, the age is inversely proportional to the Lyman continuum photon flux. If the \HII region expands in a uniform medium, the radius of an \HII region as a function of time is given by
\begin{equation}
r(t) = r_i \left( 1 + \frac{7 c_i t}{4 r_i} \right)^{4/7}
\end{equation}
where
\begin{equation}
r_i = \left( \frac{3 N_C^{'}}{4 \pi n_e n_H \beta} \right)^{1/3}
\end{equation}
is the initial Str\"omgren radius, $c_i$ is the sound speed in the ionized gas ($\sim 10\ \mathrm{km\ s}^{-1}$ for $10^4\ \mathrm{K}$), $n_e$ and $n_H$ are the ambient densities of electron and atomic hydrogen, $\beta$ is the recombination coefficient ($\sim 3 \times 10^{-13}\ \mathrm{cm}^3\ \mathrm{s}^{-1}$ for $T_e = 10^4\ \mathrm{K}$), and $N_C^{'}$ is the effective Lyman continuum photon flux \citep[e.g.][]{dyson}. $N_C^{'}$ is evaluated assuming 90 \% of the Lyman photons are absorbed by dust within the \HII region \citep{uv90}. If we adopt $n_e=n_H$, the age of an \HII region is simplified as follows:
{
\small
\setlength\arraycolsep{1pt}
\begin{eqnarray}
  t & \simeq & 0.14\ \mathrm{Myr}
    \left(\frac{n_e}{10^3\ \mathrm{cm}^{-3}}\right)^{1/2}
    \left(\frac{r}{1\ \mathrm{pc}}\right)^{7/4}
    \left(\frac{N_C^{'}}{10^{48}\ \mathrm{s}^{-1}}\right)^{-1/4} \nonumber\\
  & & -\ 0.017\ \mathrm{Myr}
    \left(\frac{n_e}{10^3\ \mathrm{cm}^{-3}}\right)^{-2/3}
    \left(\frac{N_C^{'}}{10^{48}\ \mathrm{s}^{-1}}\right)^{1/3}.
    \label{eq:age}
\end{eqnarray}}

Although the three \HII regions in W51B are not spherical, we may
roughly estimate the age of the \HII regions from equation
(\ref{eq:age}). The mean electron density of W51B \HII regions is
$\simeq 10^3\ \mathrm{cm}^{-3}$ \citep{mufson, mehringer}. The
present sizes of the \HII regions are decided from the appearance
of the 21-cm radio continuum emission (0.3 Jy beam\ind{-1} contour
in Figure \ref{13CO_H21}), and are presented in Table \ref{para}.
In the table, we also present the spectral type of the most
massive star of a cluster and the total Lyman continuum photon
flux that the cluster members emit at ZAMS stage \citep{panagia},
together with the resultant age. The Lyman flux and the inferred
age using only the stars which are expected to be already in
main-sequence are put in parentheses. We derive the age to be the
order of or younger than 1 Myr for all \HII regions. However, even
in an ideal case, the dynamic age of the \HII region may give only
the age of the most massive stars of the cluster. In addition, the
dynamic age is quite sensitive to the size of \HII region, which
depends on the distance. Nevertheless, the deduced dynamic age is
roughly consistent with the age estimation via $JHK_{\rm s}$
excess fraction, which is shown in next paragraph.

The fraction of the stars possessing NIR excess emission is
revealed to be another indicator of age \citep[and
therein]{HLLletter}. For a very young embedded cluster NGC 2024
\citep[$\sim$ 0.3 Myr,][]{meyer}, the $JHK$ excess fraction of
58\% $\pm$ 7\% is derived by \citet{HLL00}. The fraction decreases
as the age increases: 50\% $\pm$ 7\% \citep{lada00} and 21\% $\pm$
5\% \citep{HLL01} for the Trapezium with age of $\sim$ 1.5 Myr and
IC 348 with $\sim$ 2.3 Myr \citep{palla00}, respectively. From
this tendency, we can use the fractions of PMS member candidates
in order to conjecture the age of the W51B clusters. The fractions
are 66/168 (39\%), 105/589 (18\%), and 45/120 (38\%) for \G{48.9},
\G{49.0}, and \G{49.2}, respectively. Subtracting the field star
contaminations, the fractions are given by 44/108 (41\%), 39/246
(16\%), 33/89 (37\%) in order. These fractions of 20--40\% imply
that the age of the clusters are approximately 2 Myr. It should be
the upper limit of the cluster age because we detect only massive
OB stars whose NIR excess fraction is expected to be lower than
the low-mass stars due to more rapid disk dispersal times
\citep{lada00, HLL01}. However, $L$-band observation is needed for
robust estimation of cluster age because $K$-band involves risk
under the influence of nebulous environments \citep{HLL00}.

\section{SUMMARY}
We have presented results from deep \jhk band observations of two fields including three compact \HII regions along W51B GMC. We set constraints to select the possible cluster members associated with each \HII region. We investigate the IMF using $J$-band luminosity and estimate total stellar mass from a well-defined comparison IMF through the whole mass range (from $0.01 M_\odot$ to the mass of the most massive stars). In particular, we find the followings:
\begin{itemize}
\di We compute the reddening slope in the direction of W51B in the
2MASS system, by a least-squares fit with 3-$\sigma$ clipping. The
resultant color excess ratio is $E_{J-H}/E_{H-K_{\rm s}} \sim
2.07$. \di We define the cluster member candidates as the sources
within the distinct bump in the histogram along \jk~color and the
sources showing NIR color excess emission in the color-color
diagram. \di The mean visual extinctions for the cluster member
candidates are 17 mag for both \G{48.9} and \G{49.0}, and 23 mag
for \G{49.2}. \di Each IMF based on $J$-band luminosity
corresponds to the IMF of a young open cluster NGC 2264. A
universal IMF assumption provides total stellar mass of $\sim$
$1.4\times10^4 M_\odot$ , which is nearly 3 times of the observed
stellar mass. The overall SFEs are 7 \% for the cloud including
\G{48.9} and \G{49.0}, and 17 \% for \G{49.2}, which are much
higher than the Galactic average value of a few percent. \di We
estimate the dynamic age of each \HII region to be the order of or
younger than 1 Myr, in terms of the expansion of Str\"omgren
sphere of a compact \HII region. The $JHK_{\rm s}$ excess
fractions of 20--40\% imply the cluster age of $\sim$ 2 Myr, which
should be the upper limit of age in consideration of more rapid
disk dispersal times for massive stars.
\end{itemize}

\acknowledgments This work was supported by the Korea Science and
Engineering Foundation Grant R14-2002-01000-0. H.~Kim has been
partially supported by the BK21 project of the Korean Government.
H.~Sung acknowledges the support of the Korea Science and
Engineering Foundation (KOSEF) to the Astrophysical Research
Center for the Structure and Evolution of the Cosmos (ARCSEC) at
Sejong University. Finally, we are grateful to Jae-Joon Lee for
the technical assistance.

\end{document}